\documentclass[12pt]{article} 
\usepackage{graphicx} 
\def \b{{\cal B}} 
 
\def \bea{\begin{eqnarray}} 
\def \beq{\begin{equation}} 
\def \bl{\bar\lambda} 
\def \bo{B^0}

\def \cn{Collaboration} 
\def \cpp{C_{\pi \pi}} 
\def \eea{\end{eqnarray}} 
\def \eeq{\end{equation}}

\def \lpp{\lambda_{\pi \pi}} 
\def \ob{\overline{B}^0} 
\def \ok{\overline{K}^0} 
\def \rpp{R_{\pi \pi}}

\def \spp{S_{\pi \pi}} 
\begin{document} 
\begin{flushright} 
TECHNION-PH-2007-14\\ 
EFI 07-13 \\ 
arXiv:0704.3459 [hep-ph] \\ 
April 2007 \\ 
\end{flushright} 
\renewcommand{\thesection}{\Roman{section}} 
\renewcommand{\thetable}{\Roman{table}} 
\centerline{\bf SYSTEMATIC ERROR ON WEAK PHASE $\gamma$ FROM 
$B \to \pi^+ \pi^-$ AND $B \to K \pi$.
\footnote{Submitted to Physics Letters B.}} 
\medskip 
\centerline{Michael Gronau} 
\centerline{\it Physics Department, Technion -- Israel Institute of Technology} 
\centerline{\it 32000 Haifa, Israel} 
\medskip 
\centerline{Jonathan L. Rosner} 
\centerline{\it Enrico Fermi Institute and Department of Physics, University 
of Chicago} 
\centerline{\it Chicago, Illinois 60637.} 
\bigskip 
\begin{quote} 
 
When CP asymmetries in $B^0(t)\to \pi^+ \pi^-$ are combined using broken flavor 
SU(3) with decay rates for $B^+\to K^0\pi^+$ and/or $B^0\to K^+\pi^-$, one can 
obtain stringent limits on the weak phase $\gamma$ which are consistent 
with those obtained from other CKM constraints.  Experimental data in the past 
few years have improved to the extent that systematic errors associated with 
uncertainty in SU(3) symmetry breaking dominate the determination of $\gamma$.   
We obtain a value $\gamma =(73 \pm 4^{+10}_{-8})^\circ$, where the first error 
is statistical while the second one is systematic. 
 
\end{quote} 
 
\leftline{\qquad PACS codes:  12.15.Hh, 12.15.Ji, 13.25.Hw, 14.40.Nd} 
 
\begin{figure} 
\includegraphics[width=0.98\textwidth]{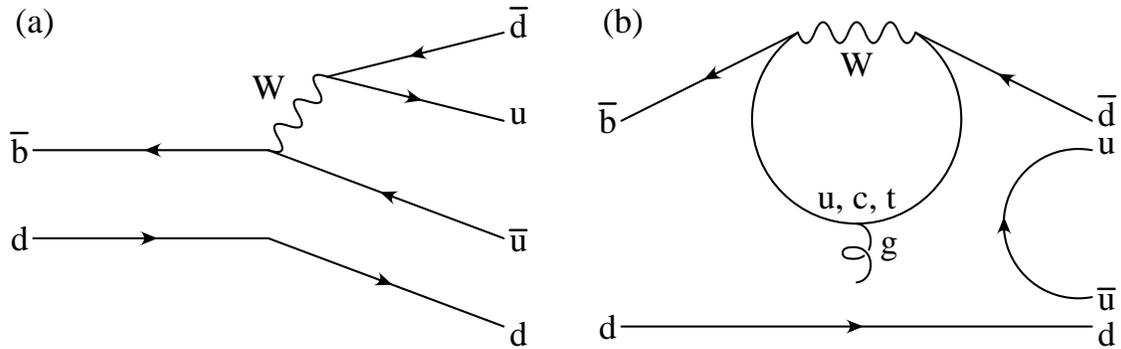} 
\caption{Examples of decay topologies for $B^0 \to \pi^+ \pi^-$.  (a) Tree; 
(b) penguin. 
\label{fig:treepen}} 
\end{figure}

\bigskip 
Time-dependent CP-violating asymmetries in the decays $B^0(t) \to \pi^+ \pi^-$ 
and their charge conjugates can provide useful information 
on the weak phase $\alpha$ or $\gamma$ of the Cabibbo-Kobayashi-Maskawa  
(CKM) matrix. These quantities 
$\cpp\equiv -A_{CP}(B^0\to\pi^+\pi^-)$ and $\spp$, defined by \cite{MG} 
\beq 
\frac {\Gamma(\ob(t) \to \pi^+\pi^-) - \Gamma(\bo(t)\to\pi^+\pi^-)} 
 {\Gamma(\ob(t) \to \pi^+\pi^-) + \Gamma(\bo(t)\to\pi^+\pi^-)} = 
 -\cpp\cos(\Delta mt) + \spp\sin(\Delta mt)~, 
\eeq 
are respectively $\cpp = 0$ and $\spp = - \sin(2 \alpha)$ in the limit in 
which a single ``tree'' amplitude, as shown in Fig.\ \ref{fig:treepen} (a), 
dominates the $B^0 \to \pi^+ \pi^-$ decay.   
The two asymmetries are modified by a contribution 
from the ``penguin'' amplitude [Fig.\ \ref{fig:treepen} (b)] \cite{MG,LP}. 
The theoretically most precise method for determining $\gamma$ in the presence of 
a penguin amplitude is based on 
applying an isospin analysis to all three $B\to\pi\pi$  decay modes and their  
charge-conjugates~\cite{Gronau:1990ka}. The current precision of this method,  
limited by a sizable decay rate for $B^0\to\pi^0\pi^0$ and by a large experimental error in  
$A_{CP}(\pi^0\pi^0)$~\cite{hfag}, does not permit a 
complete construction of the two isospin triangles describing $B\to\pi\pi$ and  
$\bar B\to\pi\pi$ amplitudes. Thus, the three $B\to\pi\pi$ decay rates and corresponding direct  
CP asymmetries lead to a rather large systematic error of $\pm 16^\circ$ in $\gamma$~\cite{Gronau:2007xg}. 
 
An alternative way of treating a subsidiary penguin amplitude in $B^0\to\pi^+\pi^-$ 
is by estimating its contribution with the help of flavor SU(3) and the decays $B \to K \pi$,  
which are dominated by a $b \to s$ penguin contribution~\cite{GR02}. In this approach, 
a theoretical uncertainty in estimating the penguin amplitude is translated into  
a smaller relative error in $\gamma$ because of the subdominant nature of a penguin  
amplitude in $B^0\to\pi^+\pi^-$. This suppression of error is expected to be somewhat  
less effective here than in $B^0\to\rho^+\rho^-$, where a smaller penguin contribution  
estimated using $B^+\to K^{*0}\rho^+$ has been shown to lead to a systematic error of  
only several degrees in $\gamma$~\cite{Beneke:2006rb}, smaller than the error  
associated with an isospin analysis in $B\to\rho\rho$~\cite{Gronau:2007xg}. 
 
In Ref.\ \cite{Gronau:2004ej} we demonstrated the precision on 
$\alpha$ obtained when deriving the penguin amplitude in $B^0\to\pi^+\pi^-$  
either from $B^+ \to K^0 \pi^+$ (a pure-penguin process) or $B^0 \to K^+ \pi^-$  
(a process with a small tree contribution).  Data available in May 2004 from 
BaBar \cite{BaBar04} and Belle \cite{Abe:2004us}, 
\beq\label{BaBe-old} 
\cpp = \left\{ \begin{array}{c} -0.19 \pm 0.19 \pm 0.05~, \cr 
-0.58 \pm 0.15 \pm 0.07~,\end{array} \right. 
\spp = \left\{ \begin{array}{c} -0.40 \pm 0.22 \pm 0.03~,~~~~ 
{\rm BaBar}~, \cr 
-1.00 \pm 0.21 \pm 0.07~,~~~~ 
{\rm Belle}~,\end{array} \right. 
\eeq 
implied averages  
\beq\label{CSave-old} 
\cpp = -0.46\pm 0.13~,~~~~~\spp = -0.74\pm 016~. 
\eeq 
Using these data and extracting the penguin amplitude from $B^0 \to K^+ \pi^-$ (just slightly 
more restrictive than using $B^+ \to K^0 \pi^+$), we found that $\alpha = (103 \pm 
17)^\circ$ or $\alpha=(107\pm 13)^\circ$, depending on SU(3)-breaking  factors.   
With the current value of $\beta = (21.3 \pm 1.0)^\circ$ 
obtained from CP asymmetries dominated by the subprocess $b \to c \bar c s$ 
\cite{hfag}, this would entail $\gamma = (56 \pm 17)^\circ$ or $\gamma = (52\pm 13)^\circ$.   
We anticipated 
that reduction of the errors by a factor of two would not present difficulties. 
 
The experimental data have improved significantly in the past few years.  
Asymmetries reported recently by BaBar~\cite{BaSC} and Belle~\cite{BeSC}, 
\beq\label{BaBe-new} 
\cpp = \left\{ \begin{array}{c} -0.21 \pm 0.09 \pm 0.02~, \cr 
-0.55 \pm 0.08 \pm 0.05~,\end{array} \right. 
\spp = \left\{ \begin{array}{c} -0.60 \pm 0.11 \pm 0.03~,~~~~ 
{\rm BaBar}~, \cr 
-0.61 \pm 0.10 \pm 0.04~,~~~~ 
{\rm Belle}~,\end{array} \right. 
\eeq 
imply averages~\cite{hfag} 
\beq\label{CSave-new} 
\cpp = -0.38\pm 0.07~,~~~~~\spp = -0.61\pm 0.08~, 
\eeq 
with errors only about half the size of errors in (\ref{CSave-old}). A similar 
reduction of errors by a factor two occurred in ratios of $B\to K\pi$ to $B\to 
\pi^+\pi^-$ branching ratios affecting the extraction of $\gamma$ [see 
Eq.~(\ref{R+,0}) below.] Old and new charge-averaged branching ratios 
for these processes, in units of $10^{-6}$, are tabulated in Table I. 
 
\begin{table} 
\caption{Old and new branching ratios for $B\to\pi^+\pi^-$ and $B\to K\pi$ 
(in units of $10^{-6}$). 
\label{tab:brs}} 
\begin{center} 
\begin{tabular}{c c c c} \hline \hline 
Year & $B^0\to\pi^+\pi^-$ & $B^+\to K^0\pi^+$ & $B^0\to K^+\pi^-$ \\ \hline 
2004  & $4.6\pm 0.4$  & $21.8\pm 1.4$ & $18.2\pm 0.8$ \\ 
2007 & $5.16\pm 0.22$ &  $23.1\pm 1.0$ & $19.4\pm 0.6$ \\ 
\hline \hline 
\end{tabular} 
\end{center} 
\end{table} 
 
The purpose of the present note is to use the improved data for obtaining $\gamma$ with an  
experimental error, and to confront a systematic theoretical error in $\gamma$ related 
to patterns of flavor SU(3) symmetry breaking.  
We shall update our analysis of Ref.~\cite{Gronau:2004ej}, using patterns  
for SU(3) breaking which differ by $\pm{\cal O}(20\%)$, quote the  
associated systematic uncertainty in $\gamma$, and compare with a contemporary  
analysis~\cite{Fleischer:2007}. 
 
The reader may consult Refs.\ \cite{GR02,Gronau:2004ej} for earlier references 
and notation.  We recapitulate the main formulae for $\cpp$ and $\spp$. 
We integrate out the top-quark contribution in the $b\to d$ penguin transition 
and use unitarity of the CKM matrix.  Absorbing a $P_{tu}$ term in the tree 
amplitude $T$, one writes  
\beq\label{Apipi} 
A(B^0\to\pi^+\pi^-) = Te^{i\gamma} + P e^{i\delta}~. 
\eeq 
The tree $T$ and penguin $P$ amplitudes, which involve magnitudes of 
CKM factors, $|V^*_{ub}V_{ud}|$ and $|V^*_{cb}V_{cd}|$, are taken to be real 
and positive and the strong phase $\delta$ is taken to lie in the range 
$-\pi \le \delta \le \pi$. For $\ob \to\pi^+\pi^-$, $\gamma \to - \gamma$. 
The asymmetries $\cpp$ and $\spp$ are given by~\cite{MG} 
\beq \label{eqn:CSpipi} 
\cpp \equiv \frac{1 - |\lpp|^2}{1 + |\lpp|^2}~~, 
~~~~\spp \equiv \frac{2 {\rm Im}(\lpp)}{1 + |\lpp| ^2}~~, 
\eeq 
where 
\beq 
\lpp \equiv e^{-2i \beta} \frac{A(\ob \to \pi^+ \pi^-)} 
{A(B^0 \to \pi^+ \pi^-)}~~~. 
\eeq 
 
Substituting (\ref{Apipi}), one obtains~\cite{GR02} 
\bea\label{C} 
\cpp & = & \frac{2r\sin\delta\sin(\beta +\alpha)}{\rpp}~, 
\\ 
\label{S} 
\spp & = & \frac{\sin 2\alpha + 2r\cos\delta\sin(\beta-\alpha) -  
r^2\sin 2\beta}{\rpp}~, 
\\ 
\label{Rpp} 
\rpp & \equiv & 1 - 2r\cos\delta\cos(\beta + \alpha) + r^2~, 
\eea 
where $r \equiv P/T$ is a ratio of penguin to tree amplitudes. 
 
In the absence of a penguin amplitude ($r=0$) one has $\cpp=0,~\spp=\sin 
2\alpha$. To first order in $r$, one finds 
\bea 
\cpp & = & 2r\sin\delta\sin(\beta + \alpha) + {\cal O}(r^2)~, 
\\ 
\spp & = & \sin 2\alpha + 2r\cos\delta\sin(\beta + \alpha)\cos 2\alpha 
+ {\cal O}(r^2)~, 
\eea 
so that in the linear approximation the allowed region in the $(\spp,\cpp)$ 
plane is confined to an ellipse centered at $(\sin 2 \alpha, 0)$, 
with semi-principal axes $2[r\sin(\beta + \alpha)\cos 2\alpha]_{\rm max}$ 
and $2[r\sin(\beta + \alpha)]_{\rm max}$.  We will use the exact expressions 
(\ref{C})--(\ref{Rpp}). 
 
Given a value of $\beta$, as already measured in $B^0(t) \to J/\psi K_S$ 
\cite{hfag}, the observables $\cpp$ and $\spp$ provide two equations for 
$\alpha$ or $\gamma$, $r$, and $\delta$.  At least one additional constraint is needed to 
determine $\alpha$ or $\gamma$. 
 
The $B\to K\pi$ decay amplitudes are described in terms of primed quantities, 
$T'$ and $P'$~\cite{GHLR}.  We introduce an SU(3) breaking factor $f_K/f_\pi$ 
in tree amplitudes assuming that these amplitudes factorize 
\cite{Beneke:1999br}
[see discussion two paragraphs below Eq.~(\ref{3ranges})],
but begin by assuming an arbitrary SU(3)-breaking factor 
$\xi_P$ in determining $P'$ from $P$, as factorization is not expected to hold 
for penguin amplitudes~\cite{Ciuchini:1997hb,SCET}: 
\beq 
\label{TpPp} 
T' = \frac{f_K}{f_\pi}\frac{V^*_{ub}V_{us}}{V^*_{ub}V_{ud}}\,T 
 = \frac{f_K}{f_\pi}\bl T~,~~ 
P' = \xi_P \frac{V^*_{cb}V_{cs}}{V^*_{cb}V_{cd}}\,P 
= - \xi_P \bl^{-1}P~. 
\eeq 
Here  
\beq 
\bl \equiv \frac{\lambda}{1 - \lambda^2/2} = 0.230~. 
\eeq 
 
Contributions of amplitudes involving the spectator quark are expected to be  
suppressed by $\Lambda_{\rm QCD}/m_b$ relative to those  
considered~\cite{GHLR,SCET}. This includes exchange and penguin  
annihilation amplitudes ($E+PA$) in $B^0\to\pi^+\pi^-$ and an annihilation  
amplitude ($A$) in $B^+\to K^0\pi^+$. Evidence for small $E+PA$ is provided 
by an upper limit on $\b(B^0\to K^+K^-)$~\cite{hfag,Blok:1997yj}.   
We will neglect these contributions, but will include the effect of $A$ in the  
systematic error.  
In this approximation one may express $B\to K\pi$ amplitudes in terms of those 
contributing to $B^0 \to \pi^+\pi^-$: 
\bea\label{AKpi+} 
A(B^+ \to K^0\pi^+) & = & -\xi_P \bl^{-1}Pe^{i\delta}~, 
\\ 
\label{AKpi-} 
A(B^0 \to K^+\pi^-) & = & - \frac{f_K}{f_\pi}\bl\,Te^{i\gamma} +  
\xi_P \bl^{-1}Pe^{i\delta}~. 
\eea 
The CP asymmetry in the first process vanishes, while that of $B^0\to K^+\pi^-$ 
\beq\label{DeltaGamma} 
\Gamma(\ob \to K^-\pi^+) - \Gamma(\bo\to K^+\pi^-) = - \xi_P \frac{f_K}{f_\pi} 
[\Gamma(\ob \to \pi^+\pi^-) - \Gamma(\bo\to \pi^+\pi^-)]~. 
\eeq 
is related to the asymmetry in $B^0\to \pi^+\pi^-$~\cite{DH,GR}, 
 
Each of the two charge averaged rates $\bar\Gamma(B^+\to K^0\pi^+)\equiv 
[\Gamma(B^+\to K^0\pi^+)+\Gamma(B^-\to \ok \pi^-)]/2$ and $\bar\Gamma(B^0 \to 
K^+\pi^-)\equiv [\Gamma(\bo\to K^+\pi^-) +\Gamma(\ob\to K^-\pi^+)]/2$ provides 
an additional constraint on the three parameters $r,~\delta$ and $\alpha$.   
Normalizing these rates by  
$\bar\Gamma(B^0\to\pi^+\pi^-) \equiv [\Gamma(\bo\to \pi^+\pi^-) + \Gamma(\ob\to 
\pi^+\pi^-)]/2$, we define two ratios 
\beq 
{\cal R}_+ \equiv  \frac{\bl^2\,\bar\Gamma(B^+\to K^0\pi^+)} 
{\bar\Gamma(B^0\to \pi^+\pi^-)}~,~~ 
{\cal R}_0 \equiv \frac{\bl^2\,\bar\Gamma(B^0 \to K^+\pi^-)} 
{\bar\Gamma(B^0\to \pi^+\pi^-)}~, 
\eeq 
given by 
\bea\label{R+} 
{\cal R}_+ &=& \frac{\xi^2_Pr^2}{\rpp}~,\\ 
\label{R0} 
{\cal R}_0 &=& \frac{\xi^2_Pr^2 + 2\xi_Pr\bl'^2\cos\delta\cos(\beta+\alpha) + \bl'^4} 
{\rpp}~,~~~~\bl' \equiv \sqrt{\frac{f_K}{f_\pi}}\bl~. 
\eea 
Using branching ratios in Table I and the lifetime ratio \cite{hfag}  
$\tau(B^+)/\tau(B^0) =1.076 \pm 0.008$, one finds the following values for ${\cal R}_+$ and 
${\cal R}_0$, 
\beq\label{R+,0} 
{\cal R}_+ = 0.220 \pm 0.013~,~~~~{\cal R}_0 = 0.199 \pm 0.010~, 
\eeq 
As mentioned, the 5\% errors here are half those quoted in Ref.\ \cite{Gronau:2004ej}. 
Here as in Ref.\ \cite{GR02} we have applied small corrections for phase 
space factors. 
 
\begin{figure} 
\begin{center} 
\includegraphics[height=3.73in]{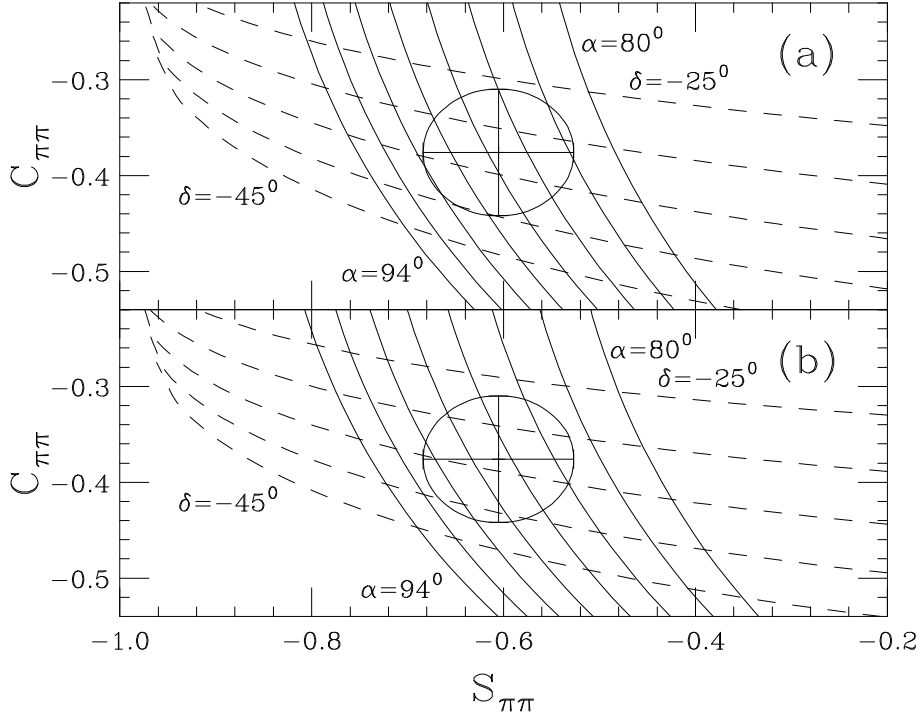} 
\end{center} 
\caption{Values of $C_{\pi \pi}$ plotted against $S_{\pi \pi}$ for values of 
$\alpha$ spaced by 2 degrees (solid curves) and $\delta$ spaced by 5 degrees 
(dashed contours), with a parameter $\xi_P = 1$ describing the degree of 
SU(3) violation in the ratio $P'/P$.  The degree of penguin ``pollution'' is 
estimated in (a) from $B^+ \to K^0 \pi^+$ and in (b) from $B^0 \to K^+ \pi^-$. 
\label{fig:cs}} 
\end{figure} 
\begin{figure} 
\begin{center} 
\includegraphics[height=3.73in]{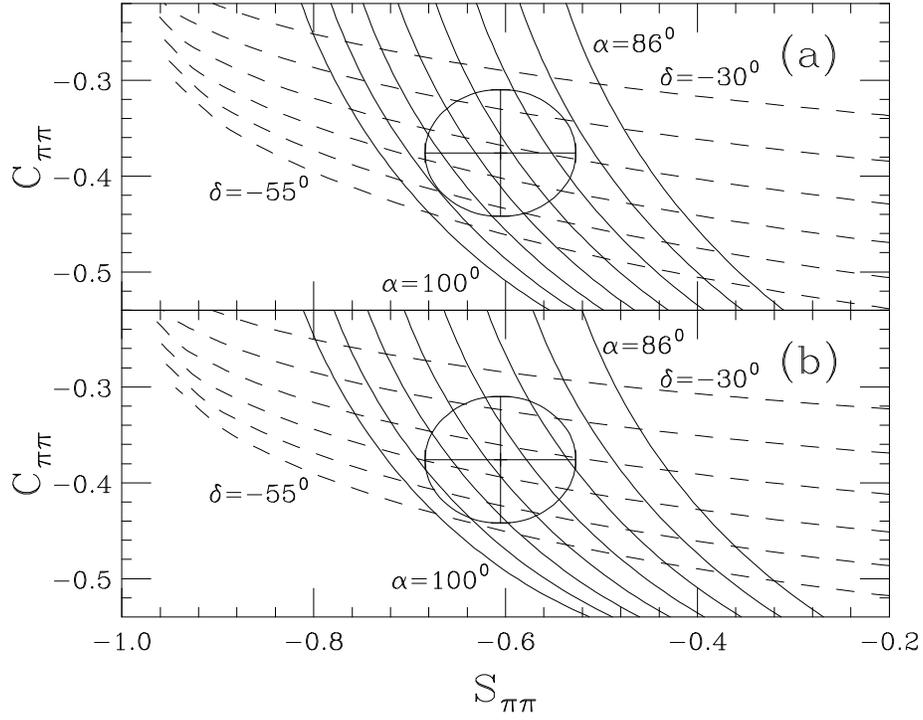} 
\end{center} 
\caption{Same as Fig.\ \ref{fig:cs} but with $\xi_P = f_K/f_\pi = 1.22$. 
\label{fig:cs+}} 
\end{figure} 
\begin{figure} 
\begin{center} 
\includegraphics[height=3.73in]{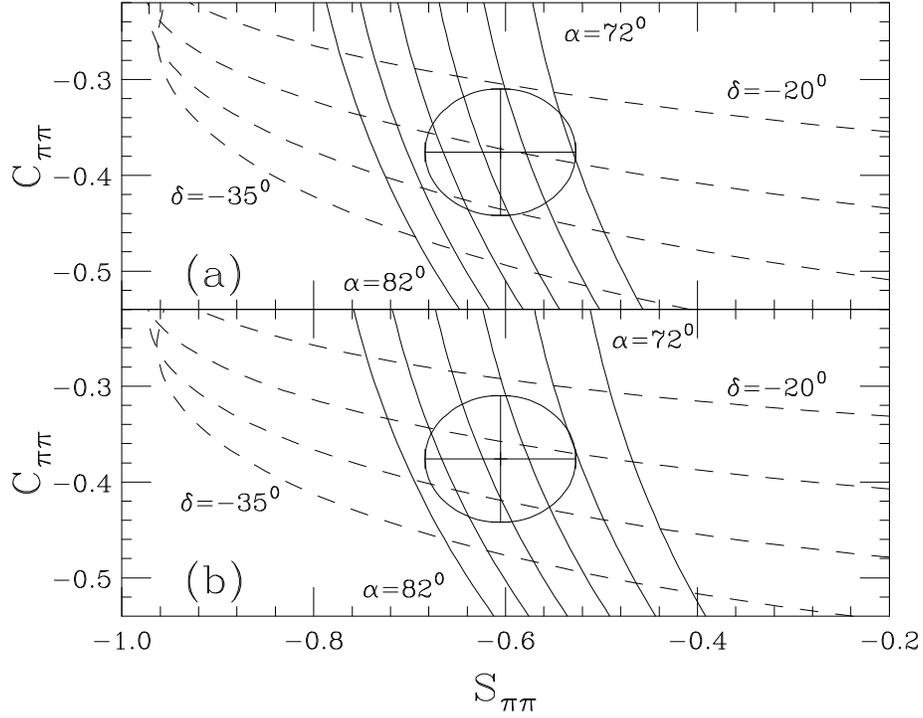} 
\end{center} 
\caption{Same as Fig.\ \ref{fig:cs} but with $\xi_P = 0.79$. 
\label{fig:cs-}} 
\end{figure} 
 
Eqs.~(\ref{C})-(\ref{Rpp}) and either (\ref{R+}) or (\ref{R0}) provide three 
equations for $r, \alpha$ and $\delta$, for given $\beta$ and for a given 
SU(3)-breaking parameter $\xi_P$ describing the ratio of $\Delta S=1$ and 
$\Delta S=0$ penguin amplitudes.  Eq.~(\ref{R+}) or (\ref{R0}) may be used to 
eliminate $r$. Thus, in Figs.\ \ref{fig:cs}, \ref{fig:cs+} and \ref{fig:cs-} we 
plot $\cpp$ and $\spp$ as functions of $\alpha$ and $\delta$ for three 
values of the SU(3) breaking parameter, $\xi_P=1,~\xi_P=f_K/f_\pi=1.22$, and 
$\xi_P=0.79$.  The latter is the central value of $\xi_P = 0.79 \pm 0.18$, 
obtained  by solving Eq.~(\ref{DeltaGamma}) using $B^0\to\pi^+\pi^-$ and 
$B^0\to K^+\pi^-$ branching ratios from Table I, the value of $\cpp$ in 
(\ref{CSave-new}), and $A_{CP}(B^0\to K^+ \pi^-) = -0.097 \pm 
0.012$~\cite{hfag}.  
 
The error ellipses in Figs.\,\ref{fig:cs}, \ref{fig:cs+} and \ref{fig:cs-} 
describing the measurements (\ref{CSave-new}) encompass somewhat different  
ranges for $\alpha$ (or $\gamma$) and $\delta$. The three corresponding pairs 
of ranges are 
\beq\label{3ranges} 
\begin{array}{c} {\rm Fig.}\,\ref{fig:cs}~(\xi_P = 1) \cr 
81^\circ \le\alpha\le 91^\circ~ \cr 
(68^\circ\le\gamma\le 78^\circ) \cr 
-42^\circ\le\delta\le-26^\circ \end{array}~~~ 
\begin{array}{c} {\rm Fig.}\,\ref{fig:cs+}~(\xi_P = 1.22) \cr 
88^\circ \le\alpha\le 99^\circ \cr 
(60^\circ\le\gamma\le 71^\circ) \cr 
-54^\circ\le\delta\le-32^\circ \end{array}~~~ 
\begin{array}{c} {\rm Fig.}\,\ref{fig:cs-}~(\xi_P = 0.79) \cr 
72^\circ \le\alpha\le 81^\circ \cr 
(78^\circ\le\gamma\le 87^\circ) \cr 
-32^\circ\le\delta\le-20^\circ~.\end{array}  
\eeq 
Here we have taken in each figure the union of the regions allowed by 
constraints (a) and (b) from ${\cal R}_+$ and ${\cal R}_0$ rather than their 
intersection.  The small differences  between the values 
following from these two constraints, at the level of a degree or two, should 
be included in the systematic rather than statistical errors. These differences 
may be associated with neglecting an annihilation amplitude in the ratio 
${\cal R}_+$. 
 
As in Ref.\ \cite{Fleischer:2007}, we find a very small statistical error in 
$\gamma$ of only 4 degrees. The systematic error in $\gamma$ associated with 
uncertainty in SU(3) breaking is larger. The change from $\xi_P= 1$ to $\xi_P = 
1.22$ and $\xi_P=0.79$ shifts $\gamma$ down by $8^\circ$ and up by $10^\circ$, 
respectively. Similarly, under these changes $\delta$ becomes more negative by 
about $10^\circ$ and less negative by about $8^\circ$, respectively. 
 
We now discuss some additional possible sources of systematic error.  They 
all indicate that the range we quote for systematic errors is probably 
conservative. 
 
(1) Because we have absorbed a $P_{tu}$ term in the tree amplitude $T$, as
noted above Eq.~(6), one might question the applicability of factorization to 
the estimate (\ref{TpPp}) of $T'/T = \bl f_K/f_\pi$.  We have investigated the 
effect of replacing $f_K/f_\pi$ in this expression by a parameter $\xi_T$ with 
range similar to that allowed for $\xi_P$.  We find very little dependence 
on $\xi_T$, with variations between 0.79 and 1.22 leading to variations of 
$\alpha$ and $\delta$ of at most a degree or two.  This may be seen from 
Eq.~(\ref{R0}) with $\bl'^2$ replaced by $\xi_T \bl^2$.  The second term in 
the numerator, proportional to $r$ and $\xi_T$, is much smaller than the 
first, proportional to $r^2$ and independent of $\xi_T$.  For a reasonable 
value of $r \sim 0.4$--0.5 and for $90^\circ < \beta+\alpha < 120^\circ$, one 
has $r^2 \sim 0.2$ while $2 r \xi_T \bl^2 |\cos(\beta + \alpha)| < 0.03$.  The 
third term in the numerator, $\xi_T^2 \bl^4$, is negligible. 
 
(2) The determinations of $\alpha$ and $\delta$ in which the penguin 
pollution in $B^0 \to \pi^+ \pi^-$ is obtained from the decay $B^+ \to 
K^0 \pi^+$ via Eq.\ (\ref{R+}) are trivially independent of $\xi_T$, as they 
do not require the estimate of $T'$ at all.  It is then reassuring that 
they are consistent within a degree or two with those obtained from 
Eq.\ (\ref{R0}). 
 
(3) The relation (\ref{DeltaGamma}) between partial width differences now 
becomes 
\beq\label{DeltaGammaXiT} 
\Gamma(\ob \to K^-\pi^+) - \Gamma(\bo\to K^+\pi^-) = - \xi_P \xi_T 
[\Gamma(\ob \to \pi^+\pi^-) - \Gamma(\bo\to \pi^+\pi^-)]~ 
\eeq 
and with the observed values of branching ratios and CP asymmetries may be 
used to constrain the product 
\beq \label{xipxit} 
\xi_P \xi_T = 0.96 \pm 0.18~ 
\eeq 
Indeed, the case illustrated in Fig.\ \ref{fig:cs+}, discussed in Ref.\ 
\cite{Fleischer:2007}, violates these bounds, and is only viable if, as in 
that work, one favors the BaBar result \cite{BaSC} implying a smaller 
direct asymmetry in $B^0 \to \pi^+ \pi^-$. 
 
\begin{figure} 
\begin{center} 
\includegraphics[height=4in]{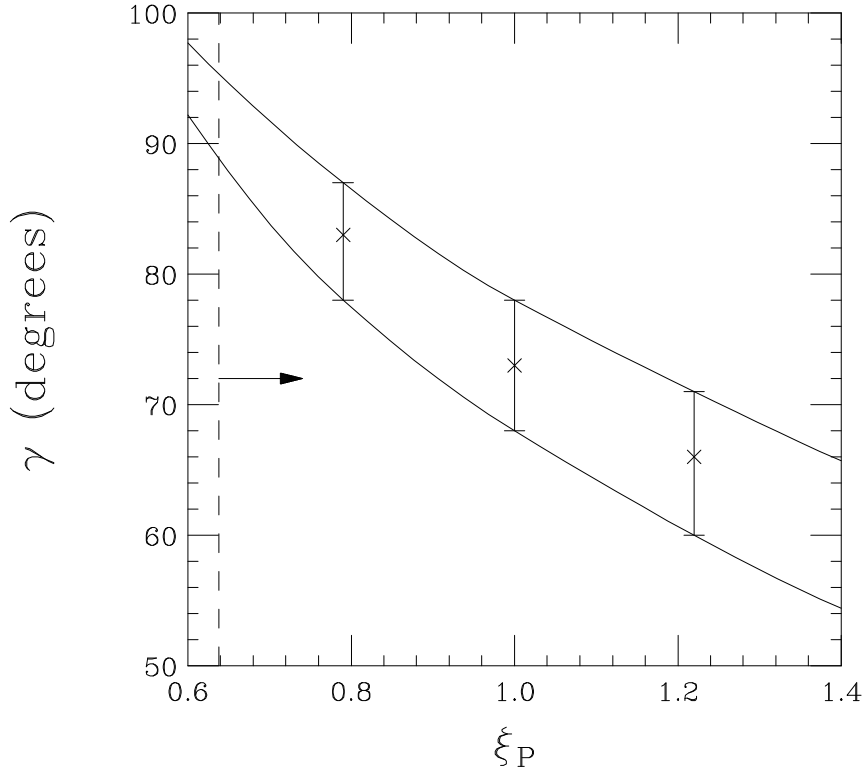} 
\caption{Dependence on $\xi_P$ of upper and lower limits on $\gamma$.  The 
three cases discussed in Figs.\ 2--4 are shown as plotted points.  The 
dashed vertical line corresponds to the lower limit $\xi_P \ge 0.64$ 
discussed in the text. 
\label{fig:xipdep}} 
\end{center} 
\end{figure}

(4) One can extrapolate beyond the values of $\xi_P$ shown in Figs.\ 2--4 
if desired.  The upper and lower limits on $\gamma$ are shown for a range 
of $\xi_P$ and the nominal value $\xi_T = 1.22$ in Fig.\ \ref{fig:xipdep}. 
The lower limit $\xi_P \ge 0.64$ is based on the $1 \sigma$ lower limit of the 
constraint $\xi_P \xi_T = 0.96 \pm 0.18$ for $\xi_T = 1.22$.  It implies only 
the rather weak bound $\gamma \le 95^\circ$.  However, even the choice 
$\xi_P = 0.79$ would suggest that SU(3) breaking acts in opposite ways for 
the tree and penguin, an unlikely circumstance given the tendencies of 
form factors involving final-state strange quarks to be enhanced relative to 
those involving nonstrange quarks. 
 
(5) The neglect of possible $E$ and $PA$ contributions is an approximation 
based on theoretical estimates which can only be fully justified once 
the branching ratio for $B^0 \to K^+ K^-$ has been shown to lie definitively 
below $10^{-7}$, as we have emphasized in several previous references 
(see, e.g., \cite{Blok:1997yj}).  The present upper limit is about three 
times this value \cite{hfag}.  One should not take the unexpectedly large 
branching ratio for $B^0 \to \pi^0 \pi^0$ as evidence for large $E + PA$, 
as it can be explained by a larger-than-expected contribution from the 
color-suppressed tree amplitude $C$ \cite{Chiang:2004nm}.

To summarize, the time-dependent asymmetries in $B^0 \to \pi^+ \pi^-$ have 
realized their statistical potential in pinning down weak phases, implying 
$\alpha = (86 \pm 4^{+8}_{-10})^\circ,~\gamma =(73 \pm 4^{+10}_{-8})^\circ$. 
The relative strong phase $\delta$ between penguin and tree amplitudes is found 
to be $\delta = (-33 \pm 7 ^{+8}_{-10})^\circ$.  The systematic errors quoted 
here are those associated with likely uncertainties in flavor-SU(3) breaking. 
Under exceptional circumstances (such as an anomalously small non-strange 
penguin amplitude) the systematic errors could even exceed those quoted. 
In order to add useful information to this 
largely model-independent discussion, explicit theoretical calculations such 
as QCD-factorization~\cite{Beneke:1999br}, Soft Collinear Effective Theory 
(SCET) \cite{SCET} or Perturbative QCD (pQCD) \cite{pQCD} need to predict 
$\delta$ with an accuracy better than the systematic error of approximately 
$\pm 10^\circ$ found above. 
 
J. L. R. is grateful to J. Tr{\^a}n Thanh V{\^a}n for the opportunity 
to present a preliminary version of these results at the XLII Rencontre 
de Moriond, La Thuile, 2007, and for hospitality during the conference.  This 
work was supported in part by the United States Department of Energy through 
Grant No.\ DE FG02 90ER40560, by the Israel Science Foundation 
under Grant No.\ 1052/04, and by the German-Israeli Foundation under 
Grant No.\ I-781-55.14/2003.

\end{document}